\documentclass[a4paper,11pt]{article}
\usepackage{jheppub} 
\usepackage{todonotes}

\newcommand{\insertfig}[2]{\includegraphics[width=#1cm]{#2}}
\newcommand \vev [1] {\langle{#1}\rangle}

\newcommand \ket [1] {|{#1}\rangle}
\newcommand \bra [1] {\langle {#1}|}
\newcommand{\ft}[2]{{\textstyle\frac{#1}{#2}}}

\title{\boldmath Five W-boson amplitude = near-null decagon}

\author[a]{A.V.~Belitsky,}
\author[b]{L.V.~Bork,}
\author[c,d]{R.N.~Lee,}
\author[d]{A.I.~Onishchenko,}
\author[e]{V.A.~Smirnov}
\affiliation[a] {Department of Physics, Arizona State University,  Tempe, AZ 85287-1504, USA}  
\affiliation[b]{The Center for Fundamental and Applied Research, 127030 Moscow, Russia}
\affiliation[c]{Budker Institute of Nuclear Physics,
630090 Novosibirsk, Russia}
\affiliation[d]{Joint Institute for Nuclear Research, 141980 Dubna, Russia}
\affiliation[e]{Skobeltsyn Institute of Nuclear Physics, Moscow State University 119992 Moscow, Russia}

\abstract{We study a five-leg scattering amplitude on the special Coulomb branch of planar $\mathcal{N} = 4$ super Yang-Mills theory. We reach this point of the moduli space of scalar vacuum expectation values by considering six-dimensional $\mathcal{N} = (1,1)$ super Yang-Mills theory and reducing it down to four space-time dimensions with extra-dimensional momenta being nonvanishing. This branch is characterized by massive external W-bosons and massless internal gluons propagating in loops. We analyze the five W-boson amplitude in the kinematics when their masses are much smaller than all Mandelstam-like invariants. This is what we dub the near mass-shell limit. We perform calculations to two-loop order in 't Hooft coupling, making use of recent advances in analytic calculations of required Feynman integrals. Our findings confirm exponentiation of infrared logarithms and enable us to conjecture a concise all-order expression for the amplitude in question. We further analyze its duality to the `square root' of a five-point correlation function of infinitely-heavy half-BPS operators, known as the decagon. By considering the near-null limit for inter-operators distances, we verify that the two objects coincide. This observation corroborates the novel Coulomb amplitudes/heavy correlator duality previously observed for four W-boson amplitudes and Sudakov form factors.}

\begin{document}
\maketitle
\flushbottom
\section{Introduction}

The predictive power of QCD dwells on two foundational pillars: asymptotic freedom \cite{Gross:1973id,Politzer:1973fx} and factorization theorems \cite{Collins:1989gx,Agarwal:2021ais}. They do not hold the same weight, though. The former is a fundamental property of the underlying non-Abelian Yang-Mills theory with unbroken SU$(N_c)$ gauge symmetry. The second one is process-dependent. Unequivocally, however, these are the latter that allow one to reduce considerations of strongly-interacting dynamics to the underlying perturbative parton scattering of quarks and gluons. This makes perturbative QCD a computationally viable theory. In the majority of applications, the partons are taken to be on their mass shell, as these were propagating over finite distances as free particles. This is counterintuitive since the quarks and gluons are not observed at asymptotically large distances: they are confined to the interior of protons/neutrons. But within their volumes, they would be detected as free had we had a chance to fit a tiny detector inside hadrons. It is the role of the asymptotic freedom to ensure the correctness of this description. How does this change as the partons go off-shell? There is a large class of physical observables where this is the case, especially when transverse components of partons' momenta become relevant. These partons develop virtualities, and their short-distance scattering occurs in the off-shell regime. How does the mathematical description change as a function of these? Traditional folklore instructed us that nothing spectacular happens in this situation. The state of affairs is nowhere near this.

How does one even ask this question in a robust manner? Naively taking partons off their mass shell breaks gauge invariance of underlying scattering processes, invalidating respective considerations from the start. Putting QCD on the back burner for now, let us address this issue within the context of a simpler Yang-Mills theory, planar $\mathcal{N} = 4$ super-Yang-Mills (sYM), where one has far more control of ab initio calculations. The model in question is quite unassuming since it possesses neither asymptotic freedom nor confinement. It depends only on a single parameter, the 't Hooft coupling $\lambda = g_{\scriptscriptstyle\rm YM}^2 N_c$ in the multicolor $N_c \to \infty$ limit. However, perturbatively, it shares a lot of properties with QCD. Historically, four-dimensional $\mathcal{N} = 4$ sYM was discovered by means of dimensional reduction of $\mathcal{N} = 1$ sYM in ten dimensions by compactifying the extra six dimensions on the torus $T^6$ \cite{Brink:1976bc,Gliozzi:1976qd}. Setting the momenta reciprocal to these extra dimensions to vanish, one obtains a theory with massless degrees of freedom. It is conformal classically and even quantum mechanically \cite{Mandelstam:1982cb}. This sYM took center stage of theoretical studies over the past thirty years, largely due to its nature of being a string theory in the disguise of Maldacena's dual pair \cite{Maldacena:1997re}. Were we to keep extra-dimensional momentum components nonvanishing, we would uncover a model with massive degrees of freedom from the four-dimensional point of view \cite{Selivanov:1999ie,Boels:2010mj}. This is the so-called Coulomb branch\footnote{It is known as such rather than the Higgs branch since the adjoint scalars of the gauge multiplet rather than hypermultiplet are endowed with vacuum expectation values.} of the theory. Without any reference to a higher-dimensional perspective, it can be generated by spontaneous breaking of gauge symmetry with the Higgs mechanism, i.e., giving vacuum expectation values (vevs) to some of the colored scalars populating the sYM Lagrangian \cite{Plefka:2014fta,Kiermaier:2011cr,Craig:2011ws}. Due to the special structure of the scalar-field potential, nonzero vevs will not break $\mathcal{N}=4$ supersymmetry. Though the R-symmetry of the theory will be deformed \cite{Craig:2011ws} and the conformal symmetry will be broken, but this will be done in a somewhat `harmless' manner. The ultraviolet properties of the theory remain unchanged, such that, for instance, the beta-function of the theory will still vanish, thus inheriting all simplifying properties of its conformal sibling residing at the origin of the moduli space of scalar vevs. Of particular interest to this work is the setup where only external lines in perturbative scattering amplitudes are massive, while the other ones, forming quantum loops, are massless. The external mass can be interpreted as an off-shellness, as we can freely take vevs to be complex. This is the point of view we will adopt in this paper.  We will call these, nevertheless as W-boson amplitudes, referring to their massive nature, rather than off-shell gluons. 

The aforementioned ten-dimensional progenitor does not enjoy an unconstrained spi\-nor-helicity formalism, and this hampers its immediate use for efficient calculation of loops. However, we can do dimensional reduction in a stepwise manner, first, to massless six-dimensional $\mathcal{N} = (1,1)$ sYM with unbroken SU($N_c$) gauge group, and then, to massive $\mathcal{N} = 4$ sYM. The `intermediate' $\mathcal{N} = (1,1)$ model does possess an unconstrained spinor-helicity formalism  \cite{Cheung:2009dc,Dennen:2009vk,Bern:2010qa}, which can be employed in tandem with the unitarity-cut sewing technique \cite{Bern:1994zx,Bern:1994cg,Bern:2004cz} to build bases of integrals at each loop order for multileg scattering amplitudes \cite{Bern:2010qa,Brandhuber:2010mm,Bork:2013wga,Belitsky:2024rwv}. This is the most fruitful route to analyze the off-shell scattering.

Having established the proper framework to address these amplitudes, we would like to uncover properties, if any, that they share with their massless counterparts. The latter possess (i) universal all-order infrared properties \cite{Sen:1982bt,Sterman:2002qn,Aybat:2006mz,Dixon:2008gr,Becher:2009qa}, (ii) an intriguing iterative perturbative structure \cite{Anastasiou:2003kj,Bern:2005iz}, (iii) a hidden relation to null polygonal Wilson loops \cite{Alday:2007hr,Drummond:2007cf,Brandhuber:2007yx}, (iv) a connection to the light-cone limit of correlators of light half-BPS operators \cite{Alday:2010zy}, and last but not least, (v) they are amenable to an integrability framework in the multi-collinear limits \cite{Basso:2013vsa}. Do the off-shell amplitudes echo any of these? In this paper, we elaborate on counterparts of (i) and (iv).

Off-shell amplitudes in planar $\mathcal{N} = 4$ sYM are finite quantities: they are free both from ultraviolet and infrared divergences alike. Recent years have taught us that as they are brought infinitesimally close to their mass shells, the transition to the origin of the moduli space is non-analytic in the off-shellness parameter, and their singularity structure differs from their massless siblings at the strict origin. While the Sudakov double logarithms of the latter are governed by the ubiquitous cusp anomalous dimension, those of the former are driven by a completely different function of 't Hooft coupling: the so-called octagon anomalous dimension. This was confirmed in a number of studies of amplitudes \cite{Caron-Huot:2021usw,Bork:2022vat} and form factors \cite{Belitsky:2022itf,Belitsky:2023ssv,Belitsky:2024agy,Belitsky:2024dcf} as well. But what about infrared-safe finite parts? These are the only remainders that contribute to physical cross sections after cancellations due to the Kinoshita-Lee-Nauenberg theorem. This question is addressed in this paper for the five-leg amplitude, building on recent advances in multiloop calculations of off-shell Feynman integrals near their mass shell \cite{Belitsky:2025sin,Bork:2025ztu}. We unravel a beautiful all-order form of this amplitude. At the same time, we also uncover its dual description in terms of a correlation function of infinitely-heavy half-BPS operators \cite{Coronado:2018ypq}. This generalizes an earlier observation tied to four-point amplitudes.

Our subsequent presentation is organized as follows. In the next section, we recall the dually-conformal covariant form of tree scattering amplitudes of 
$\mathcal{N} = (1,1)$ sYM, their use for the construction of the two-loop integrand of the five-gluon amplitude, and their further dimensional reduction to 
four space-time dimensions, landing us on the Coulomb branch in question. In Sect.\ \ref{ConformalSection}, we review infrared properties of amplitudes on the
conformal branch. Then, in Sect.\ \ref{CoulombSection}, we present our main result for the five W-boson amplitude in the near mass-shell limit: the two-loop
results and a conjecture for its all-order form. Finally, in Sect.\ \ref{DualitySection}, we explore its relation to the so-called decagon, i.e., the `square-root' of a 
five-point correlator of half-BPS operators. Closing this paper, we conclude and point out future directions of research. An appendix collects concise, explicit 
results for all Feynman integrals we encountered in our calculation.

\section{Coulomb branch from generalized dimensional reduction}

The main infrared-sensitive `observable' addressed in this work is the amplitude of five massive W-bosons. As we pointed out in the Introduction, it will be obtained from massless six-dimensional $\mathcal{N} = (1,1)$ sYM with the unbroken gauge group. The construction proceeds as follows. All of the theory's on-shell states are classified 
according to the little group $\mbox{SU}(2) \times \mbox{SU}(2)$ of its Lorentz group $\mbox{SO}(5,1) \simeq \mbox{SU}^\ast(4)$ 
\cite{Cheung:2009dc,Dennen:2009vk}. The $R$-symmetry of the model is $\mbox{SU}_{R} (2) \times \mbox{SU}_{R} (2)$, however, one sacrifices it in favor of the little group such that only its $\mbox{U}_{R} (1) \times \mbox{U}_{R} (1)$ subgroup is left manifest. This truncation allows one to accommodate all propagating degrees of freedom 
\begin{align*}
&\mbox{gluons:} &&  g^a{}_{\dot{a}} \, , \\
&\mbox{scalars:} &&  \phi, \phi^\prime, \phi^{\prime\prime}, \phi^{\prime\prime\prime}  \, , \\
&\mbox{gluinos:} &&  \chi^a, \bar\chi_{\dot{a}}, \psi^a, \bar\psi_{\dot{a}}  \, ,
\end{align*}
in a single CPT self-conjugate non-chiral superfield \cite{Dennen:2009vk}
\begin{align}
\label{6Dsuperfield}
{\mit\Phi} = \phi + \chi^a \eta_a + \bar{\chi}_{\dot{a}} \bar\eta^{\dot{a}} + \eta^2 \phi^\prime + \bar\eta^2 \phi^{\prime\prime}
+
g^{a}{}_{\dot{a}} \eta_a \bar\eta^{\dot{a}} 
+ 
\psi^a \eta_a \bar\eta^2
+
\bar\psi_{\dot{a}} \bar\eta^{\dot{a}} \eta^2
+
\eta^2 \bar\eta^2 \phi^{\prime\prime\prime}
\, ,
\end{align}
as a terminating expansion in the independent Grassmann variables $\eta_a$ and $\bar\eta_{\dot{a}}$ that carry only the 
little group indices and possess positive and negative chirality, respectively.

Scattering amplitudes in the theory are generated from the amputated vacuum expectation value of a product of these superfields, schematically
\begin{align}
\label{6Damplitude}
\mathcal{A}_n 
&  
=
\vev{{\mit\Phi}_1 \dots {\mit\Phi}_n} 
\, ,
\end{align}
such that their expansion in $\eta/\bar\eta$ generates all component amplitudes. They depend on $n$ bosonic momenta $P_i$ and $n+n$ of (anti-)chiral charges $Q_i$ and $\bar{Q}_i$, cumulatively called supermomenta. The theory benefits from a spinor-helicity formalism \cite{Cheung:2009dc}, which allows one to recast the 
super-Poincar\'e quantum numbers in terms of unconstrained Weyl spinors  \cite{Cheung:2009dc} $\Lambda_i^{A, a} \equiv \ket{i^{a}} = \bra{i^{a}}$ 
and $\bar\Lambda_{i, A, \dot{a}} \equiv |i_{\dot{a}}] = [i_{\dot{a}}|$,
\begin{align}
\label{SpinorHelicityRep}
P_i = \ket{i^{a}} \bra{i_a}
\, , \qquad
\bar{P}_i = |i_{\dot{a}}] [i^{\dot{a}}|
\, , \qquad
Q_i
=
\bra{i^{a}} \eta_{i, a}
\, , \qquad
\bar{Q}_i
=
[i_{\dot{a}}| \bar\eta_i^{\dot{a}}
\, .
\end{align}
Imposing super-momentum conservation, we can extract it in terms of bosonic and fermionic delta functions \cite{Dennen:2009vk}
\begin{align}
\mathcal{A}_n 
=
i (2 \pi)^6 
\delta^{(6)} \left( \sum\nolimits_{i = 1}^n P_i \right) 
\delta^{(4)} \left( \sum\nolimits_{i = 1}^n Q_i \right)
\delta^{(4)} \left( \sum\nolimits_{i = 1}^n \bar{Q}_i \right)
\widehat{\mathcal{A}}_n 
\, ,
\end{align} 
and define reduced amplitudes $\widehat{\mathcal{A}}_n$ which are homogeneous polynomials of order $n - 4$ in the Grassmann
variables. The non-chiral nature of the theory imposes a more stringent constraint on these: they can be chosen to be polynomials
of equal degrees $[n/2] - 2$ both in the chiral and anti-chiral charges \cite{Plefka:2014fta}, yielding natural reduction properties to
four dimensions. It is straightforward to extract the amplitude of six-dimensional gauge bosons by integrating $\mathcal{A}_n$ over the
Grassmann variables of individual legs with a proper weight that saturates all $\eta$'s,
\begin{align}
[A_n]_{a_1 a_2 \dots a_n}^{{\dot a}_1 {\dot a}_2 \dots {\dot a}_n} = \int \prod_{i = 1}^n d^2 \eta_i d^2 \bar\eta_i \, \eta_{i, a_i} \bar\eta_i^{{\dot a}_i} \mathcal{A}_n
\, .
\end{align}
Throughout this work, we will tacitly assume that the little group indices are chosen in a manner corresponding to the four-dimensional MHV amplitudes $A_n (--+\dots+)$, i.e., $a_1 = a_2 = 1$, ${\dot a}_1 = {\dot a}_2 = 1$ and all the rest $a_i = 2$,  ${\dot a}_i = 1$ ($i \geq 3$). From now on, we do not display these explicitly.

\subsection{Trees}

The four- and five-leg color-ordered reduced tree amplitudes are particularly compact when written in proper variables, the so-called {\sl dual} variables. They admit the form \cite{Cheung:2009dc,Dennen:2009vk,Plefka:2014fta}
\begin{align}
\label{CompactA4}
\widehat{\mathcal{A}}_4^{\rm tree} 
&
=
\frac{1}{S_{12} S_{23}}
\, ,\\
\label{CompactA5}
\widehat{\mathcal{A}}_5^{\rm tree} 
&
=
\frac{- \Omega_{12345}}{S_{12} S_{23} S_{34} S_{45} S_{51}}
\, .
\end{align}
The denominators exhibit the fact that these develop only two-particle poles in the Mandelstam-like variable $S_{ij} = (P_i+P_j)^2$ as their momenta `coalesce'. The numerator 
\begin{align}
\label{MinimalElements}
\Omega_{ijklm} = \ft12  \bra{B_{i,jl}} \bar{B}_{i,km}] + {\rm cc}
\, ,
\end{align}
is determined by the ${\rm SU}^{\ast}(4)$-invariant inner product of bras and chiral-conjugate\footnote{The cc-operation is defined by changing all unbarred symbols to barred ones and vice versa.} (cc) kets,
\begin{align}
\label{BconfCovs}
\bra{B_{i,jk}} \equiv \bra{{\mit\Theta}_{ij}} \bar{X}_{jk} X_{ki} + \bra{{\mit\Theta}_{ik}} \bar{X}_{kj} X_{ji}
\, , \qquad
|\bar{B}_{i,jk}] \equiv - {\rm cc} \big( \bra{B_{i,jk}} \big)
\, .
\end{align} 
Here, for the dual variables $X$, ${\mit\Theta}$, and $\bar{\mit\Theta}$, 
\begin{align}
\label{superPoincare2Dual}
P_i = X_{i, i+1}
\, , \qquad
Q_i = {\mit\Theta}_{i, i+1}
\, , \qquad
\bar{Q}_i = \bar{\mit\Theta}_{i, i+1}
\, ,
\end{align}
we use the convention 
\begin{align}
X_{ij} \equiv X_i - X_j 
\, , \qquad
{\mit\Theta}_{ij} \equiv {\mit\Theta}_i - {\mit\Theta}_j
\, , \qquad
\bar{\mit\Theta}_{ij} \equiv \bar{\mit\Theta}_i - \bar{\mit\Theta}_j
\, .
\end{align}
Under the conformal inversion\footnote{In these equations, we temporarily restored explicit SU$^\ast$(4) Lorentz indices.}, 
\begin{align}
\label{Inversions}
\mathcal{I} X^{AB} = \frac{\bar{X}_{AB}}{X^2}
\, , \qquad
\mathcal{I} {\mit\Theta}^A = {\mit\Theta}^B \frac{\bar{X}_{BA}}{X^2} 
\, , \qquad
\mathcal{I} \bar{\mit\Theta}_A = \frac{X^{AB}}{X^2} \bar{\mit\Theta}_B 
\, ,
\end{align}
the amplitude transforms covariantly \cite{Dennen:2010dh,Huang:2011um,Plefka:2014fta}
\begin{align}
\label{ConformalInversionAhat}
\mathcal{I} \widehat{\mathcal A}^{(0)}_n = X_1^2 \dots X_n^2 \widehat{\mathcal A}^{(0)}_n
\, .
\end{align}
This property is very important as it will be inherited by loop integrals to be discussed next.

\subsection{Loops and dimensional uplift}

The above spinor-helicity formalism allows one to build loop integrands by means of the unitarity-cut sewing procedure \cite{Bern:1994zx,Bern:1994cg,Bern:2004cz}. This is by far the most efficient modern way to construct gauge-invariant expressions for amplitudes in terms of scalar Feynman integrals, bypassing explicit use of heavily redundant, gauge-dependent Feynman rules. Normalizing the all-order amplitude $A_{n}$ to its tree, 
\begin{align}
\label{M4DimD}
A_{n} = A_n^{\rm tree} M_n
\end{align}
the ratio function $M_n$ of Mandelstam-like invariants $S_{ij}$ develops the perturbative expansion 
\begin{align}
M_n = 1 + \lambda_6 \, M_n^{(1)} + \lambda_6^2 \, M_n^{(2)}+ \ldots
\, ,
\end{align}
in terms of the dimensionful six-dimensional 't Hooft coupling 
\begin{align}
\label{6Dcoupling}
\lambda_6 = g_{6, \scriptscriptstyle\rm YM}^2 N_c
\, .
\end{align}

To date, the four-leg amplitude has received by far the most attention. It was found to the staggering four-loop order \cite{Bern:2010qa} from its cuts into gauge-invariant trees (\ref{CompactA4}-\ref{CompactA5}). For instance, at one loop, it is given by the sum of two $\sigma_2$-cyclic permutations of the box integral
\begin{align}
\label{1loopExamp}
M_4^{(1)}
=
\frac{i}{2} \sum_{\sigma_2} \int \frac{d^6 K}{(2\pi)^6} \frac{S_{12} S_{23}}{K^2 S_{1K} S_{12K} S_{123K}}
\, .
\end{align} 
It has the very same form as in four dimensions, except that now all Lorentz-invariant products 
\begin{align}
\label{6DMandelstamDef}
S_{ij\dots K} \equiv (P_i + P_j  \dots + K)^2
\end{align}
are six-dimensional. The most profound lesson that we learn from this consideration is that planar loop integrands can be deduced from their four-dimensional counterparts $s_{ij \dots k}= (p_i + p_j  \dots + k)^2$ by a dimensional uplift,
\begin{align}
s_{ij \dots k}
\to 
S_{ij \dots K}
\, .
\end{align}
Another important conclusion is that integrands inherit the six-dimensional dual conformal covariance of tree amplitudes.  

With a little ingenuity \cite{Belitsky:2024rwv}, the five-leg amplitude can also be easily found in a fully covariant six-dimensional form to the lowest two perturbative orders. The one-loop integrand is again determined by (now, five cyclic permutations of) the above box
\begin{align}
\label{M5OneLoop}
M_5^{(1)}
=
\frac{i}{2} \sum_{\sigma_5} \int \frac{d^6 K}{(2\pi)^6} \frac{S_{12} S_{23}}{K^2 S_{1K} S_{12K} S_{123K}},
\end{align} 
while the two-loop ratio function
\begin{align}
\label{M5TwoLoop}
M_5^{(2)}
&
=
\frac{i^2}{2} \sum_{\sigma_5} \int \frac{d^6 K_1}{(2\pi)^6}\frac{d^6 K_2}{(2\pi)^6} 
\bigg[
\frac{S_{12} S_{23} S_{45} S_{4, -K_1}}{K_1^2 K_2^2 S_{K_1 K_2} S_{3K_1} S_{23 K_1} S_{123 K_1} S_{4 K_2} S_{45 K_2}}
\\
&
+
\frac{S_{34} S_{45}^2}{K_1^2 K_2^2 S_{K_1 K_2} S_{3K_1} S_{123 K_1} S_{4 K_2} S_{45 K_2}}
+
\frac{S_{51} S_{45}^2}{K_1^2 K_2^2 S_{K_1 K_2} S_{23K_1} S_{123 K_1} S_{4 K_2} S_{45 K_2}}
\bigg]
\, , \nonumber
\end{align}
is given in terms of the sum of a pentabox (first line) and two double box (second line) integrals. Here, the integrands are written in terms of the Lorentz invariant products introduced in Eq.\ (\ref{6DMandelstamDef}), e.g., $S_{K_1 K_2} = (K_1+K_2)^2$, etc.  The pentabox possesses a loop-momentum-dependent irreducible numerator to ensure its proper transformation properties under the dual conformal inversion (\ref{Inversions}) to match the ones of the double boxes. Again, as for the four-leg case, this set of integrals is exactly the same as in four dimensions. It agrees with a ten-dimensional analysis of the five-point integrands in Refs.\ \cite{Mafra:2008ar,Mafra:2015mja}. 

The found simple pattern of dimensional uplift will undoubtedly persist for higher-leg amplitudes, provided one is able to extract loop integrands from the unitarity cuts in a six-dimensional Lorentz-covariant manner. They will also inherit corresponding dual-conformal covariant properties of the tree building blocks. Then their generalized dimensional reduction to the special Coulomb branch in four dimensions will be insensitive to the so-called $\mu$-terms\footnote{The above statement of preserving six-dimensional covariance is instrumental for this as the $\mu$-terms arise from the extra-dimensional $6-4$ momentum components, see, e.g., Ref.\ \cite{Bern:2010qa}.} \cite{Bern:2002tk,Bern:2008ap}, which do break good dual inversion properties of integrands.

\subsection{Generalized dimensional reduction}
\label{GDRsection}

Now, we are in a position to demonstrate how we pass to the special Coulomb branch of the theory, where only external legs are kept off-shell. We will begin, however, with quite a general setup where there are as many masses as there are external/internal lines. To remain comprehensive, we illustrate this with a simple example of the one-loop four-leg ratio function $M_4^{(1)}$. The best way to introduce masses in a manner consistent with dual conformal symmetry is to transform it first to the dual variables introduced in Eq.\ (\ref{superPoincare2Dual}). It is important to realize that this is possible only in the planar limit. The on-shell condition $P_i^2=0$ for external momenta in these dual coordinates will be translated to the null condition $P_i^2=X_{i,i+1}^2=0$ for the interval $X_{i,i+1}$. Let us split the six-dimensional coordinates $X_i = (x_i, y_i)$ into the four-/extra-dimensional components $x_i/y_i$. Then we compactify the extra dimensions on the equal $R_0$-radii $T^2$-torus. Sending $R_0 \to 0$, we can implement this limit as the following constraint on the integration measure
\begin{align}
d^6X_0 \mapsto (2\pi R_0)^2 d^6X_0 \delta^{(2)}(y_0)
\, .
\end{align}
This will give us consecutively
\begin{align}
\label{1loopExamp1}
M_4^{(1)}
&
=\frac{i}{2} \sum_{\sigma_2}
\int \frac{d^6 X_0}{(2\pi)^6} \frac{X_{13}^2X_{24}^2}{X_{10}^2X_{20}^2X_{30}^2X_{40}^2}
\nonumber\\
&
\mapsto
\frac{i}{2} R_0^2 \sum_{\sigma_2} \int \frac{d^4 x_0}{(2\pi)^4} \frac{(x_{13}^2-y_{13}^2)(x_{24}^2-y_{24}^2)}{(x_{10}^2-y_1^2)(x_{20}^2-y_2^2)(x_{30}^2-y_3^2)(x_{40}^2-y_4^2)}
\, .
\end{align}
We observe that squares of out-of-four-dimensional components $y_i^2$ can be interpreted from the $D=4$ point of view as masses $M_i^2$ of virtual states propagating in loops, while $y_{i,i+1}^2$ as the squared masses $m_{i}^2$ of external particles\footnote{In principle, the masses $m_{i}$ (not their squares!) of external particles obey additional linear constraints due to the $6$-dimensional momentum conservation \cite{Craig:2011ws}, but this will be irrelevant for our discussion.} 
\begin{align}
M_i^2 \equiv y_i^2 
\, , \qquad
p_i^2 = m_{i}^2 \equiv y_{i,i+1}^2
\, .
\end{align}
The latter stem from the six-dimensional on-shell condition $P_i^2 = X_{i,i+1}^2 = x_{i,i+1}^2 - y_{i,i+1}^2 = p_i^2-m^2_i=0$, with $p_i$ being the $D=4$ momenta of external legs. From the four-dimensional point of view, these masses can be produced by a choice of vevs for the model's scalars \cite{Craig:2011ws,Caron-Huot:2021usw}. For generic values of $M_i$ and $m_i$, the integral (\ref{1loopExamp1}) is finite and can be evaluated directly in $D=4$ without an additional regularization. This consideration immediately generalizes to any number of loops and legs, provided we have a basis set of scalar Feynman integrals to work with. 

The focus of the current consideration is on a small vicinity of the  moduli space near its origin, 
\begin{align}
M_i^2=0 \, , \qquad 
m_i^2 = m^2 \to 0
\end{align}
for all $i$. This corresponds to the regime when all propagators in loop integrals are massless, while the single mass parameter of all external legs is non-zero but infinitesimally small\footnote{The opposite regime where all $m_i^2=0$ and all $M_i^2=M^2\ll1$ was considered in a series of papers and is interesting in its own right \cite{HennGiggs1,Henn:2010ir,Henn:2011by}. See discussion and comparison between these two regimes in \cite{Caron-Huot:2021usw,Bork:2022vat}.} compared to generically-valued Mandelstam-like invariants $s_{ij\dots}$. We will think about external square masses as being negative $m^2 < 0$ so that they are viewed as external legs' virtualities. We will dub this kinematical situation as the {\sl near mass-shell} regime. The limit of small $m \to 0$ is obviously singular, and corresponding infrared divergences will now manifest themselves as powers of the logarithm $\log m$.

From now on, we will absorb the power of the radius $R_0$ together with $1/(4 \pi)^2$ from the loop integral measure into the definition of the four-dimensional 't Hooft coupling
\begin{align}
g^2 = \frac{\lambda_6 R_0^2}{(4 \pi)^2} 
\equiv
\frac{g_{\scriptscriptstyle\rm YM}^2 N_c}{(4 \pi)^2}
\, , 
\end{align}
with $\lambda_6$ from Eq.\ (\ref{6Dcoupling}). With this normalization, the per-loop integration measure becomes 
\begin{align}
\label{IntMeasure}
\int_k \equiv i \int \frac{d^4 k}{\pi^2}
\, .
\end{align}
Using these conventions, the perturbative series for the ratio functions reads 
\begin{align}
M_n = 1 + \sum_{\ell \geq 1} g^{2 \ell} M_n^{(\ell)}
\, ,
\end{align}
where, for instance, for four legs at one loop, passing back to the momentum space, we have
\begin{align}
\label{1loopExamp2}
M_4^{(1)}
=
\frac{1}{2} \sum_{\sigma_2} \int_k \frac{s_{12} s_{23}}{k^2 s_{1k} s_{12k} s_{123k}}+ O(m^2)
\, .
\end{align}
This four-dimensional integral with massless propagators and external virtual legs is the well-known Davydychev-Usyukina box function \cite{Usyukina:1992jd}.

\section{Iterative structure of conformal branch amplitudes}
\label{ConformalSection}

Before diving into perturbative properties of amplitudes in the near mass-shell regime of the Coulomb branch of $\mathcal{N}=4$ sYM, it is instructive to recall the situation at the strict origin, i.e., purely massless on-shell setup. At this conformal point, there are no intrinsic scales and thus momentum integrals are divergent. While they do not possess ultraviolet singularities since the model is finite, infrared singularities do arise whenever loop momenta are soft or become collinear to external light-like momenta. These infrared divergences require a regulator. Dimensional regularization with $D = 4 - 2\varepsilon$, or rather its version in the form of dimensional reduction that preserves supersymmetry, is chosen for this purpose such that singularities arise as poles in $\varepsilon$. Studies of infrared poles in massless amplitudes in gauge theories go back to the early eighties of the last century and stretch into the modern day \cite{Mueller:1979ih,Magnea:1990zb,Sterman:2002qn,Bern:2005iz}. They established the following picture. In the planar limit, infrared divergences are accompanied by dependence on Mandelstam invariants involving only adjacent legs, $s_{i,i+1}$. This is an immediate reflection of the fact that as their momenta become collinear, i.e.,  $p_i \to z (p_i + p_{i+1})$ and $p_{i+1} \to (1-z) (p_i + p_{i+1})$, this dependence is entirely encoded in the on-shell two-particle Sudakov form factor $\mathcal{F}^{\rm on}_2$,
\begin{align}
\label{AmplFactor}
\log M_n = \frac{1}{2}\sum_{i=1}^n\log \mathcal{F}_2^{\rm on}  \left(\frac{\mu^2}{s_{i,i+1}}; g,\varepsilon\right) 
+
r_n \big(s,g\big) 
+
O(\varepsilon)
,
\end{align}
while the remainder function $r_n$ is infrared-safe as $\varepsilon \to 0$ and depends on the 't Hooft coupling as well as two- and more-particle Mandelstam invariants, here, cumulatively called by $s$. The on-shell Sudakov form factor, per se, was found to exponentiate infrared poles order-by-order of the perturbative series. It admits the following form \cite{Mueller:1979ih,Magnea:1990zb,Sterman:2002qn} 
\begin{align}
\label{FF2}
\log \mathcal{F}^{\rm on}_2 \left(\frac{\mu^2}{s}; g,\varepsilon\right)
=
-
\frac{1}{2}\sum_{\ell=1}^{\infty} g^{2 \ell}
\left[ \frac{\Gamma^{(\ell)}_{\rm cusp}}{(\ell\varepsilon)^2}
+
\frac{G^{(\ell)}}{(\ell\varepsilon)}
+
c^{(\ell)}\right]
\left(- \frac{\mu^2}{s}\right)^{\ell \varepsilon}
+
O(\varepsilon)
\, .
\end{align}
The right-hand side is determined in terms of two infinite sets of transcendental numbers  $\Gamma^{(\ell)}_{\rm cusp}$ and $G^{(\ell)}$, known as coefficients of the cusp $\Gamma_{\rm cusp} (g)$ and collinear $G (g)$ anomalous dimensions. They read in the first two orders of perturbation theory \cite{Polyakov:1980ca,Korchemsky:1987wg,Anastasiou:2003kj,Bern:2005iz}
\begin{align}
\label{GCuspAndGPT}
\Gamma_{\rm cusp}(g)
=
\sum_{\ell = 1}^{\infty} \Gamma_{\rm cusp}^{(\ell)} g^{2\ell}
=
4 g^2 - 8 \zeta_2 g^4 + \ldots
\, , \quad
G(g)
=
\sum_{\ell = 1}^{\infty} G^{(\ell)} g^{2\ell}
=
-
4 \zeta_3 g^4 
+
\ldots
\, .
\end{align}

The finite remainders, as they appear in Eq.\ (\ref{AmplFactor}), are constrained neither by arguments related to the existence of infrared singularities nor by the breaking of the scaling and special conformal boosts due to the emergence of the dimensionful scale $\mu$ from the regularization procedure. However, the violation of dual symmetries, in particular of the dual inversion, is highly consequential. Its well-understood pattern of breaking was formulated in terms of anomalous Ward identities which unambiguously predicted the functional dependence of $r_4$ and $r_5$ on Mandelstam-like variables to all orders in 't Hooft coupling. The solution to these was found to be determined by the cusp anomalous dimension \cite{Drummond:2007au,Drummond:2008aq}
\begin{align}
\label{FinPartDimReg}
r_4 (s, g)
&
=+
\frac{\Gamma_{\rm cusp}(g)}{4}
\log^2 \frac{s_{12}}{s_{23}}
+
c_4(g),\nonumber\\
r_5 (s, g)
&
=
-
\frac{\Gamma_{\rm cusp}(g)}{8}\sum_{i=1}^5
\log \frac{s_{i,i+1}}{s_{i+1,i+2}} 
\log \frac{s_{i-1,i}}{s_{i+2,i+3}}
+
c_5(g).
\end{align}
Here $c_{4,5}(g)$ are kinematically-independent functions of 't Hooft coupling, which are obviously unconstrained by the Ward identities. 

Why are the latter nevertheless so powerful? The reason is that they take the form of inhomogeneous linear differential equations in dual variables $x_i$. For the four- and five-leg case, one cannot form conformal cross-ratios which are invariant under conformal boosts and, thus, would be oblivious to the inhomogeneity in these equations \cite{Drummond:2007au}. Thus, once the inhomogeneity is fixed by the cusp anomaly $\Gamma_{\rm cusp}(g)$, the corresponding equations completely define unique solutions in Mandelstam invariants. For six or more legs, these Ward identities do not suffice, as there are nontrivial cross ratios in these observables. But it is a different story.

For $n=4,5$, there is an equivalent way to present scattering amplitudes by means of the so-called BDS ansatz \cite{Anastasiou:2003kj,Bern:2005iz,Bern:2006vw},
\begin{align}
\label{BDSExpDimReg}
\log M_n^{\rm BDS}
=
\sum_{\ell=1}^{\infty}g^{2L}
\big[\Gamma_{\rm cusp}^{(\ell)} + \ell \varepsilon G^{(\ell)} + \varepsilon^2 C^{(\ell)} + \dots \big] M_n^{(1)} (\ell\varepsilon)
\, .
\end{align}
It displays a profound iterative structure of the perturbative series on the conformal branch of $\mathcal{N} = 4$ sYM and provides a prediction for higher orders in terms of the lowest-loop amplitude and a set of transcendental numbers. Do Coulomb branch amplitudes, in particular, the four- and five-leg ones, share any of these properties? This is what we turn to next.

\section{Special Coulomb branch amplitudes}
\label{CoulombSection}

Let us now turn back to the main character of our play: the amplitudes of W-bosons in the near-mass-shell kinematical regime. On general grounds of universality of infrared divergences, one anticipates that these scattering amplitudes should obey a factorization akin to their conformal counterparts
\cite{Bork:2022vat,Belitsky:2022itf,Belitsky:2023ssv}:
\begin{align}
\label{AmplFactorOffShell1}
\log M_n = \frac{1}{2}\sum_{i=1}^n
\log \mathcal{F}_{2}^{\rm off}\left(\frac{m^2}{s_{i,i+1}};g\right) + f_n\big(s,g\big)+\frac{n}{2} D(g) + O(m^2),
\end{align}
where, however, we replaced the on-shell Sudakov form factor $\mathcal{F}_{2}^{\rm on}$ with a potentially different off-shell version $\mathcal{F}_{2}^{\rm off}$. It is determined by the same matrix element as its on-shell version, but evaluated in the near mass-shell kinematics. This factorization is structurally identical to the conformal case (\ref{AmplFactor}), except for a marginal variation in defining the finite remainder, but there is an unexpected twist in the plot. Based on explicit {\sl three-loop} calculations, one can conjecture the following all-order formula for $\mathcal{F}_{2}^{\rm off}$ \cite{Belitsky:2022itf},
\begin{align}
\label{FFFactorOffShell}
\log\mathcal{F}_{2}^{\rm off}
\left(\frac{m^2}{s};g \right) =  
-
\frac{\Gamma_{\rm oct} (g)}{2}\log^2\frac{m^2}{s}
-
D(g) + O(m^2),
\end{align}
with both functions of 't Hooft coupling known exactly \cite{Belitsky:2019fan}
\begin{align}
\label{GammaOctSeries}
\Gamma_{\rm oct}(g)
&
=\frac{2}{\pi^2}\log \cosh \left(2 \pi g\right)  = 4 g^2-16 \zeta_2 g^4 + \ldots,\nonumber\\
D(g)
&
=\frac{1}{4}\log \frac{\sinh( 4 \pi g )}{4\pi g} = 4 \zeta_2 g^2 - 32 \zeta_4 g^4 + \ldots
\, .
\end{align}
We observe from this that, while the infrared physics is indeed captured by the Sudakov form factors, its perturbative structure is different. It cannot be, for instance, reproduced by a naive replacement of infrared poles in the parameter of dimensional regularization with logarithms of the off-shellness, i.e., $1/\varepsilon \mapsto \log m^2$. In addition to the difference between the functions of the coupling,  $\Gamma_{\rm cusp}$ vs.\ $\Gamma_{\rm oct}$, for the leading logarithms, there is no trace of the collinear anomalous dimension in $\mathcal{F}_{2}^{\rm off}$ for subleading ones. These conclusions shake the foundation of the common belief that $\Gamma_{\rm cusp}$ controls infrared behavior of scattering amplitudes at all points of the moduli space in planar $\mathcal{N}=4$ sYM. It also reaffirms earlier results pointing to supersession of the former lead actor $\Gamma_{\rm cusp}$ by the understudy $\Gamma_{\rm oct}$ in other off-shell `observables' as reported in Refs.\ \cite{Caron-Huot:2021usw,Bork:2022vat,Belitsky:2022itf,Belitsky:2023ssv,Belitsky:2024agy,Belitsky:2024dcf}.
 
Before we proceed with explicit calculations of the remainder functions $f_n$, let us make a comment. While the representation (\ref{AmplFactorOffShell1}) is very natural from the point of view of emphasizing infrared physics, it is not the most optimal one for manifesting explicit and hidden symmetries that these amplitudes possess. Indeed, away from the origin of the moduli space, all loop ratio functions $M_n$ are finite and, thus, enjoy dual conformal {\sl symmetry}, not just covariance. This implies that they can be recast in terms of conformal cross-ratios since $p_i^2 = m^2 = x_{i,i+1}^2\neq 0$ in contrast to the conformal point. This is the case even for $n=4,5$. The limit $m \to 0$, although singular, can be discussed in terms of corresponding cross-ratios without any issues. Depending on their precise definition, they will merely possess certain scaling with $m$. Conditions stemming from this will be restrictive regarding the functional dependence of ratio functions. 

\subsection{Four W-bosons}

As a first case of study, let us recall the form of the amplitude $M_4$ for four W-bosons introduced in Ref.\ \cite{Caron-Huot:2021usw}. It is a function of two conformal cross ratios
\begin{align}
\label{2pCC1}
w_1=\frac{x_{12}^2x_{34}^2}{x_{13}^2x_{24}^2}
\, , \qquad
w_2=\frac{x_{23}^2x_{41}^2}{x_{13}^2x_{24}^2}
\, ,
\end{align}
both of which are zero at the origin of the moduli space. However, in our regime of $x_{i,i+1}^2 = p_i^2=m^2 \ll s_{12} \sim s_{23}$, they are small, but nonvanishing, and are equal to each other
\begin{align}
\label{2pCC2}
w \equiv w_1=w_2=\frac{m^4}{s_{12} s_{23}} \, .
\end{align}
The dual conformal invariance and the expected exponentiation of the double-logarithmic behavior (\ref{AmplFactorOffShell1}) of $M_4$ as $m \to 0$, immediately yields the prediction
\begin{align}
\label{M4dci}
\log M_4 = -\frac{\Gamma_{\rm oct}(g)}{4}\log^2 w - \frac{D(g)}{2} + O(m^2)
\, ,
\end{align}
up to an additive function of the coupling $D(g)$. The latter can be found from explicit multiloop analyses. Currently, at least up to three-loop order, everything points out that this $D (g)$ is given by the exact function of 't Hooft coupling in Eq. (\ref{GammaOctSeries}). From the above result, the remainder function for four W-bosons reads
\begin{align}
\label{f4}
f_4=\frac{\Gamma_{\rm oct}(g)}{4}\log^2 \frac{s_{12}}{s_{23}} 
-
\frac{D(g)}{2}
\, .
\end{align}

\subsection{Five W-bosons}

In this section, we present the main result of this work. Applying the generalized dimensional reduction, reviewed in Section \ref{GDRsection} to the six-di\-men\-sional five-leg amplitude, given by Eqs.\ (\ref{M5OneLoop}-\ref{M5TwoLoop}) to two-loop order, we find
\begin{equation}
\label{M5expan}
M_5 = 1+g^2 M_{5}^{(1)}+g^4 M_{5}^{(2)}+\ldots
\end{equation} 
with one- and two-loop ratio functions 
\begin{align}
\label{M5int}
M^{(1)}_5
&= 
\ft{1}{2} \mathbb{P}_5 \mbox{Box},
\nonumber\\
M^{(2)}_5 
&=
\ft{1}{2}\mathbb{P}_5\left[ \mbox{DBox}_1 +\mbox{DBox}_2  + \mbox{PBox} \right] 
\, ,
\end{align}
being defined in terms of the off-shell box\footnote{The graph assigns a scalar propagator for each internal line. Numerators in terms of Mandelstam-like invariants and irreducible scalar products are written explicitly.}
\begin{align}
\label{BoxEq}
\mbox{Box}
&
=
s_{12} s_{23}
\parbox[c][35mm][t]{45mm}{
\insertfig{4.1}{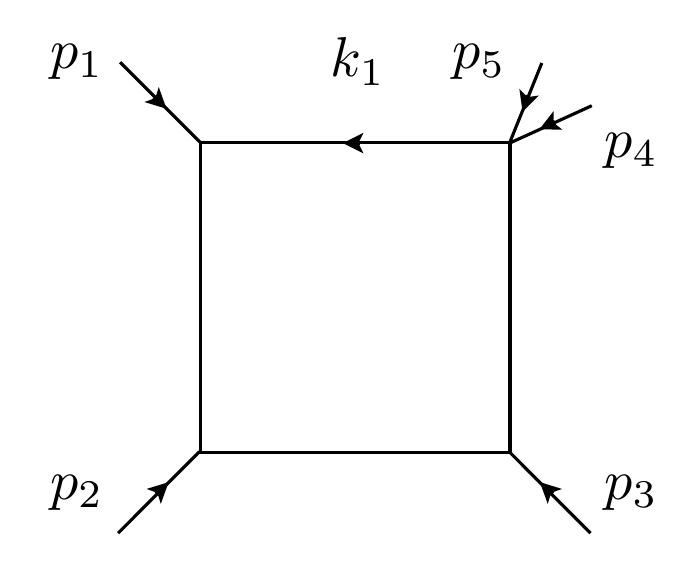}
}
\, ,
\end{align}
the off-shell double box
\begin{align}
\mbox{DBox}_1
&
=
s_{34} s_{45}^2
\ \ \,
\parbox[c][33mm][t]{58mm}{
\insertfig{5.7}{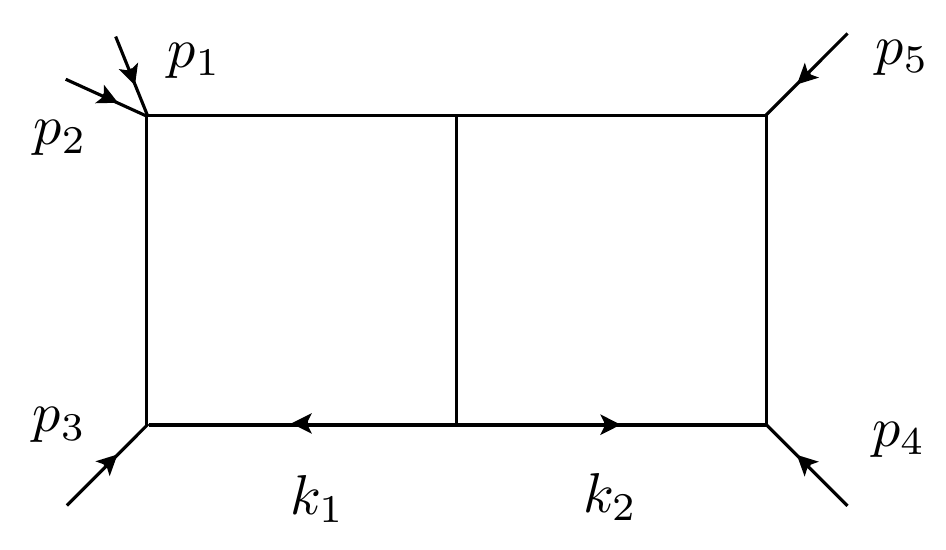}
}
\, , \\
\mbox{DBox}_2
&
=
s_{51} s_{45}^2
\, 
\parbox[c][30mm][t]{60mm}{
\insertfig{6}{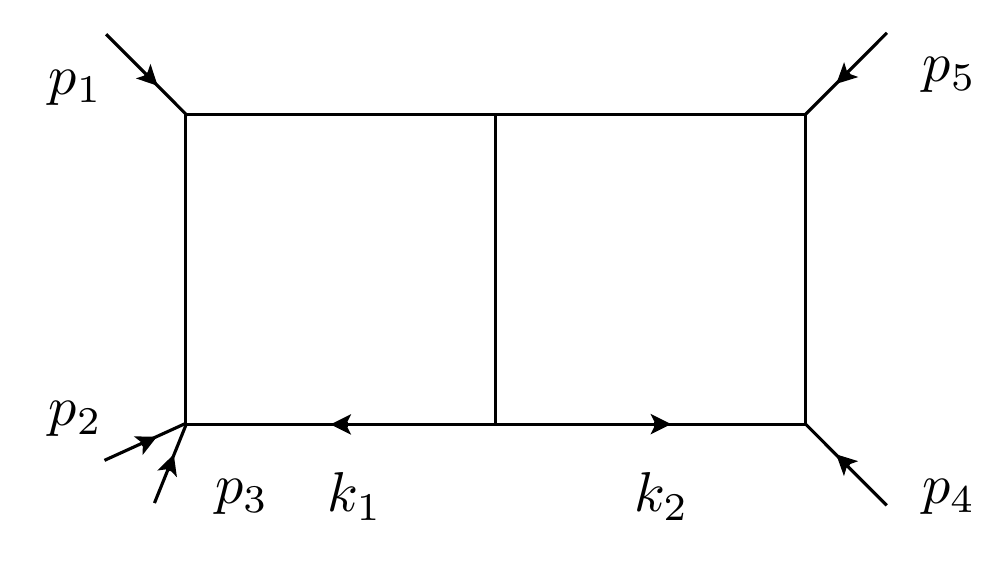}
}
\, ,
\label{DBoxEq}
\end{align}
and the off-shell pentabox 
\begin{align}
\mbox{PBox}
=
s_{12} s_{23} s_{45}
s_{4,-k_1}
\parbox[c][31mm][t]{65mm}{
\insertfig{6.5}{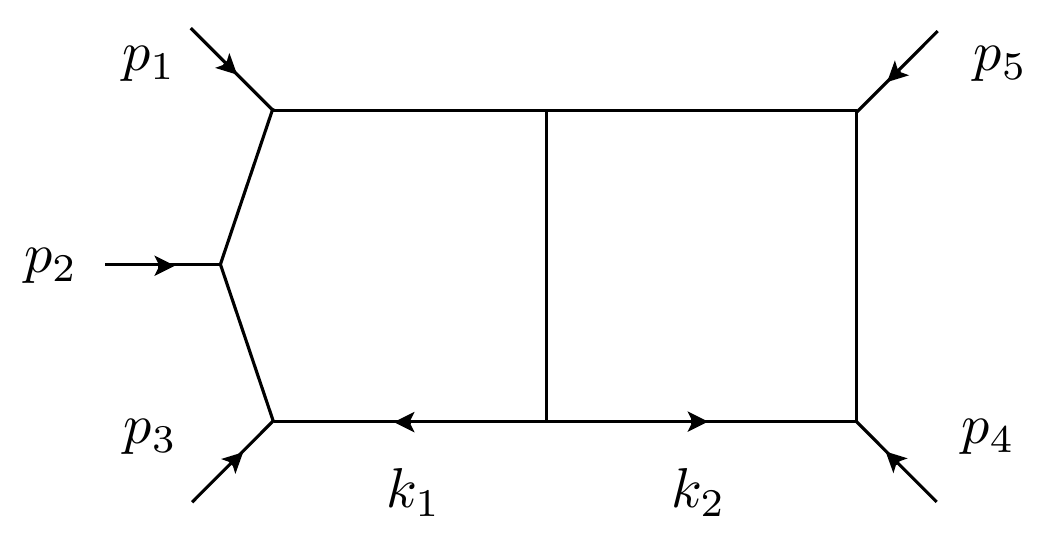}
}
\, ,
\label{PBoxEq}
\end{align}
integrals, respectively, with the per-loop integration measure defined in Eq.\ (\ref{IntMeasure}). These expressions confirm earlier considerations which relied on a naive dimensional uplift of massless scalar integrals defining $M_5$ on the conformal branch by merely endowing external legs with a mass \cite{Bork:2022vat}. The five cyclic permutations are enabled with the help of a shift operator $\mathbb{P}_5$. It is given by
\begin{equation}
\mathbb{P}_5=\sum_{n=0}^4\mathbb{P}^n=1+\mathbb{P}+\mathbb{P}^2+\mathbb{P}^3+\mathbb{P}^4,
\end{equation}
where individual terms are powers of the operators $\mathbb{P}$, which shifts leg's labels of external W-bosons by $+1$, $\mbox{mod}(5)$. Explicitly, its action reads $\mathbb{P}^n f_i = f_{i+n}$. 

The ratio $M_5$ is a function of five independent Mandelstam-like variables $s_{ii+1} = (p_i + p_{i+1})^2$ where $i = 1, \dots, 5$, and the off-shellness $m^2 = p_i^2$ (for any $i$). The functional space of the above integrals as $m \to 0$ is expected to be spanned by Goncharov polylogarithms of weight $4-p$  accompanying powers of infrared logarithms $\log^p m^2$ with $p\leqslant 4$. While the double box integrals are particular cases of the well-known Davydychev-Usyukina ladders \cite{Usyukina:1992jd,Usyukina:1993ch}, the near mass-shell pentabox was not known until a couple of months ago. Recent advances in multiloop calculations allowed us to find it in an analytic form  \cite{Belitsky:2025sin,Bork:2025ztu}. We defer very concise solutions for all of the required integrals to Appendix \ref{Appendix}.

Keeping track, at first, only of infrared-sensitive logarithms, we observe that they  exponentiate, as expected, and confirm the off-shell Sudakov form factor (\ref{FFFactorOffShell}) as the correct quantity to absorb them, 
\begin{equation}\label{M5div}
\left. \log M_5 \right|_{\rm div}= \frac{1}{2}
\mathbb{P}_5\log \mathcal{F}_{2}^{\rm off}\left(\frac{m^2}
{s_{12}};g\right)+\frac{5}{2}D(g)= - \frac{\Gamma_{\rm oct} (g)}{4}\mathbb{P}_5\log^2\frac{m^2}{s_{12}} \, .
\end{equation}
This agrees with an earlier study \cite{Bork:2022vat} at a single point in the parameter space with identical values of Mandelstam-like variables $s_{i,i+1} = Q^2$ for any $i=1,\dots,5$.

The structure of the finite part $f_5$ of the amplitude is, however, more interesting. Defining the perturbative expansion of $f_5$ as
\begin{equation}
f_5
=g^2f_5^{(1)}+g^4f_5^{(2)}+\ldots
\, ,
\end{equation}
we find the following one-loop expression
\begin{equation}
\label{FP1}
f_5^{(1)}=-\frac{1}{2}\mathbb{P}_5\log \frac{s_{12}}{s_{45}} \log \frac{s_{23}}{s_{34}} - 5\zeta_2
\, ,
\end{equation}
and a very concise result in two loops 
\begin{align}
\label{FP2}
f_5^{(2)}=\mathbb{P}_5\bigg[\frac{3}{2} \zeta_2 \log\frac{s_{12}}{s_{45}} \log\frac{s_{23}}{s_{34}}-\zeta_2\log\frac{s_{12}}{s_{23}} \log\frac{s_{12}}{s_{34}}\bigg] + \frac{135}{4} \zeta_4
\, .
\end{align}
Both of these are represented solely in terms of logarithms! This functional dependence echoes the conformal case (\ref{FinPartDimReg}), but a naked eye inspection clearly shows numerical differences in the coefficients accompanying corresponding functions.

It is instructive to recast our result for the ratio function $M_5$ in terms of conformal cross-ratios since the dual conformal invariance is exact. To this end, one can introduce two equivalent sets of variables. They differ only in their scaling with respect to the off-shellness $m$. We define these as
\begin{align}
\label{UsAndVs}
u_i = \frac{x_{i+1, i+2}^2 x_{i-1, i-2}^2}{x_{i+1, i-2}^2 x_{i+2, i-1}^2}
\, , \qquad
v_i = \frac{x_{i+1, i-1}^2 x_{i+2, i-2}^2}{x_{i+1, i-2}^2 x_{i-1, i+2}^2}
\, ,\qquad\mbox{for}\qquad
i = 1,\dots, 5
\, .
\end{align}
They are related to each other as follows
\begin{align}
u_i = v_{i-1} v_{i+1}
\, ,
\end{align}
with all labels defined $\mbox{mod(5)}$. All $u$'s and $v$'s are of order $m^4$ and $m^2$, respectively, as can be seen from their explicit expressions in our special kinematics
\begin{equation}\label{uset}
u_i = \frac{m^4}{s_{i+1,i+2} s_{i+2,i+3}},
\, ,\qquad
v_i = \frac{m^2 s_{i-1,i}}{s_{i+1,i+2} s_{i+2,i+3}}
\, .
\end{equation}
We will refer to them, correspondingly, as the $u$- and $v$-bases. Assembling our findings together, we find in the $v$-basis
\begin{align}
\label{logM5w}
\log M_5
= 
&
- \left(g^2 - 4\zeta_2 g^4 \right) \mathbb{P}_5\log v_1\log v_2
\\
&
-
g^4 \zeta_2 \mathbb{P}_5 \log \frac{v_1}{v_2} \log \frac{v_2}{v_3}
-  5\zeta_2 g^2
+
\frac{135}{4} \zeta_4 g^4 
+
O(g^6)
\, , \nonumber
\end{align}
and, equivalently, in the $u$-basis
\begin{align}
\label{logM5u}
\log M_5= 
&
- \ft14 \left( g^2 - 4 \zeta_2 g^4\right) \mathbb{P}_5\log^2 u_1
\\
&
-
\ft12
\left( g^2 - 2 \zeta_2 g^4 \right) \mathbb{P}_5 \log u_1
\log \frac{u_2}{u_3}-  5\zeta_2 g^2
+
\frac{135}{4} \zeta_4 g^4  + O(g^6)
\, . \nonumber
\end{align}
This two-loop result motivates us to suggest the following all-loop conjecture for $M_5$ 
\begin{align}
\label{logM5uAllLoop}
\log M_5
&
= 
-
\mathbb{P}_5 \left[
\frac{\Gamma_{\rm oct}(g)}{16} \log^2 u_1
+
\frac{\Gamma_{\rm cusp}^\prime (g)}{8} \log u_1 \log \frac{u_2}{u_3}
\right]
+ d_5(g)
\, , \\
\log M_5
&
= 
-
\mathbb{P}_5 \left[
\frac{\Gamma_{\rm oct}(g)}{4} \log v_1\log v_2
+
\frac{\Gamma_{\rm cusp}^\prime(g) - \Gamma_{\rm oct}(g)}{8}
\log \frac{v_1}{v_2} \log \frac{v_2}{v_3}
\right]
+ d_5(g)
\, ,
\end{align}
in the above two bases. Here, $d_5$ is a function of the 't Hooft coupling, with its first two terms in the perturbative series being
\begin{equation}\label{d5}
d_5(g)= -  5\zeta_2 g^2
+
\frac{135}{4} \zeta_4 g^4 
+
\ldots\, .
\end{equation}
Notice that the second representation above allows for a more natural breakup of the amplitude in terms of the infrared-sensitive first term and a finite remainder. Intriguingly, this form resembles the structure observed for the remainder function of the six-leg amplitude in the origin-limit \cite{Basso:2020xts} of its cross-ratios, though the functions of the coupling accompanying logrithmic terms are different. Namely, in our proposal, $\Gamma^\prime_{\rm cusp}(g) = 4 g^2 - 8 \zeta_2 g^4 + O (g^6)$ matches the lowest two perturbative orders of the ubiquitous cusp anomalous dimension. However, we lack sufficient information to confirm or refute their coincidence exactly in 't Hooft coupling. The main dilemma that we face is that, up to now, we have not found a single observable on the special Coulomb branch where coefficients accompanying log-dependent functions are expressed in terms of odd values of the Riemann zeta function. This was made evident in the absence of a counterpart for the collinear anomalous dimension in all two-loop calculations. Studies at three loops would not be sufficient to dis-/prove our conundrum, unfortunately. It would, however, be definitely conclusive regarding the rigidity of the kinematic structure in Eq.\ (\ref{logM5uAllLoop}) against perturbative effects in $g$, especially the second term there. The four loops are the first order where $\Gamma_{\rm cusp}(g)$ develops dependence on odd zetas. Currently, however, this goal is beyond our reach.

\section{Duality to heavy correlation functions}
\label{DualitySection}

\begin{figure}[t]
\begin{center}
\mbox{
\begin{picture}(0,100)(160,0)
\put(0,0){\insertfig{4.5}{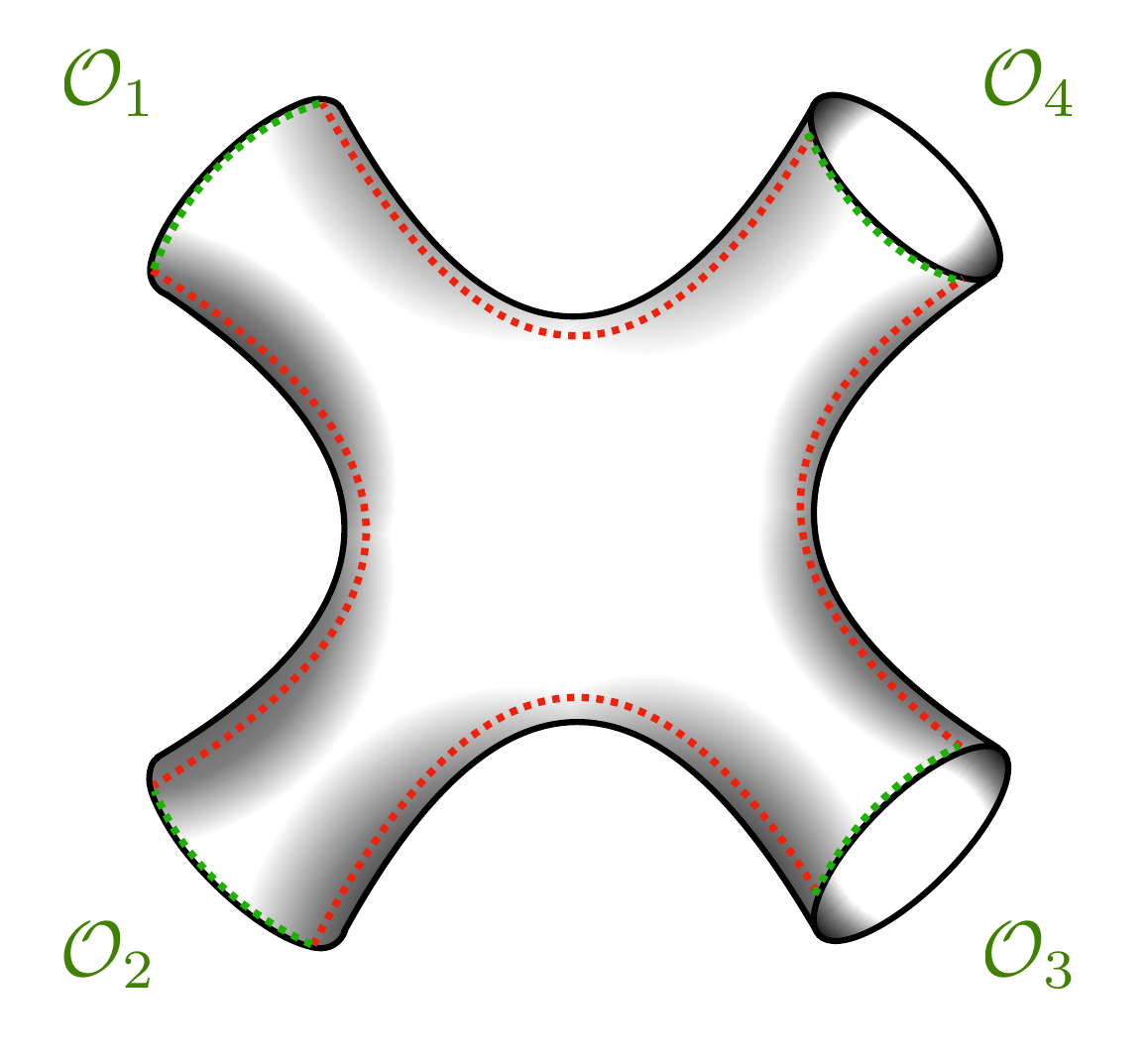}}
\put(60,56){$\mathbb{O}_0$}
\put(190,-12){\insertfig{5}{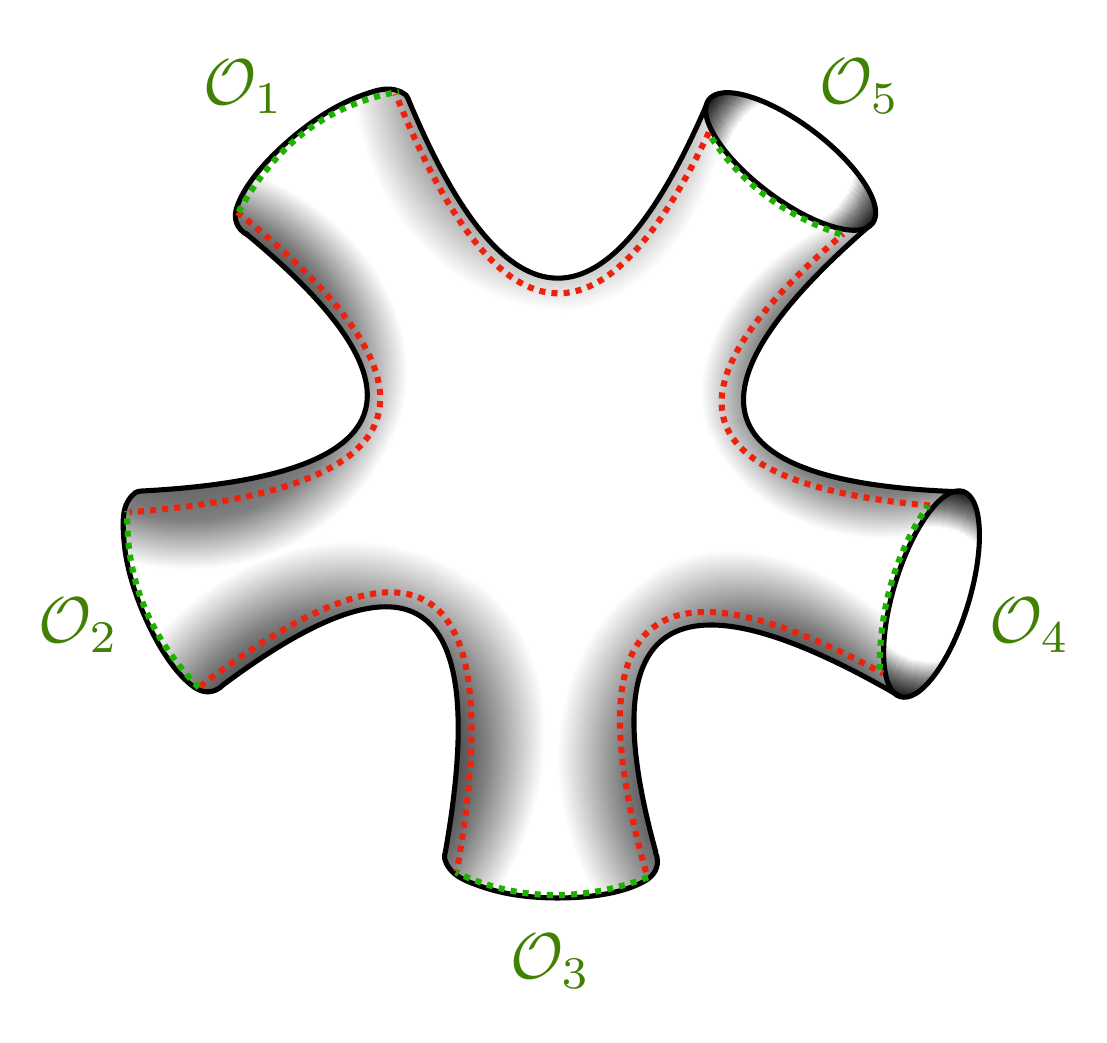}}
\put(255,56){$\mathbb{D}_0$}
\end{picture}
}
\end{center}
\caption{\label{WorldSheetPic} World-sheet representation of correlation functions of four- (left) and five-point correlation functions.}
\end{figure}

Scattering amplitudes $M_n$ on the conformal branch of $\mathcal{N} = 4$ sYM obey a very intriguing hidden relation to Wilson loops $W_{C_n}$ on null polygonal contours $C_n$ in the fundamental representation \cite{Alday:2007hr,Drummond:2007cf,Brandhuber:2007yx}. The contour's segments are determined by the gluons' momenta $p_i$ beginning/ending at cusps located at dual coordinates $x_i$, such that $p_i = x_{i, i+1}$. Its quantum field-theoretical origin is still obscure, but it naturally arises from the string-theoretical perspective as a T-duality transformation of the world-sheet of Maldacena's dual type IIB theory on AdS$_5\times$S$^5$ background \cite{Berkovits:2008ic}. In fact, one can further generalize this amplitude/Wilson loop duality to a triality \cite{Alday:2010zy}, with the third ingredient being the planar correlation function 
\begin{align}
\label{GMmasslessDual2}
G_n
=
\langle {\rm tr} Z^2 (x_1) \, {\rm tr} \bar{Z}^2 (x_2) \dots {\rm tr} \bar{Z}^2 (x_n) \rangle,
\end{align}
of lightest half-BPS operators\footnote{Here, $Z$ and $\bar{Z}$ are the complex scalar and its complex conjugate of $\mathcal{N} = 4$ sYM. Below, we will also encounter the other two, $X$ and $Y$.} in the pair-wise light-cone limit \cite{Alday:2010zy},
\begin{align}
\label{GMmasslessDual1}
\lim_{x_{i,i+1}^2 \to 0} 
G_n/G_n^{{\rm tree}} 
= 
M_n^2
\, .
\end{align}
Its origin is more transparent since in the null limit $x_{i,i+1}^2 \to 0$, the propagator stretching between any two adjacent operators is dominated by long-wavelength gluon emissions, which factorize in the leading twist approximation into an adjoint Wilson line on the segment $[i,i+1]$ accompanying a free-particle propagator. Assembling these together for all pairs in $G_n$, and taking the planar limit, allows one to `split' the adjoint loop into the square of the fundamental ones. This establishes then the three-fold way symmetry between $M_n$, $W_{C_n}$ and $\sqrt{G_n}$.

Is there a similar triality for Coulomb branch amplitudes? Yes, there is, at least, a duality! The pair in question involves correlation functions of infinitely-heavy half-BPS operators, instead.

\subsection{Octagon}

The four-point correlator of half-BPS operators with a very large R-charge $K$ \cite{Coronado:2018ypq,Coronado:2018cxj}
\begin{equation}\label{Korr}
G_{4, K} =\langle \mbox{tr}\bar{X}^{2K}(x_1)\ \mbox{tr}X^K\bar{Z}^K(x_2)\ \mbox{tr}Z^{2K}(x_3)\ \mbox{tr}X^K\bar{Z}^K(x_4) \rangle
\, ,
\end{equation}
was dubbed the {\sl simplest} correlation function \cite{Coronado:2018ypq} for a reason. For finite values of $K$, but in the planar limit, gauge interactions fill out a genus-zero world-sheet in the color space with four punctures where the single-trace operators are inserted. It is demonstrated in the left panel of Fig.\ \ref{WorldSheetPic}. For virtual gluons to cross the `bridges' formed by the free propagators stretched between nearest-neighbor operator pairs (shown by red dashed curves in Fig.~\ref{WorldSheetPic}), one has to go to the $K$-th perturbative order in $g^2$. Therefore, for sufficiently large values of $K$, gluons cannot pass from one side of the world-sheet to another. The front and back completely decouple such that \cite{Coronado:2018ypq,Coronado:2018cxj}
\begin{align*}
\lim_{K \to \infty} G_{4, K}/G_{4, K}^{\rm tree} = \mathbb{O}^2 (w_1,w_2; g)
\, ,
\end{align*}
the correlation function is given by the square of the so-called octagon $\mathbb{O}$ (formed by connecting red and green dashed lines in Fig.\ \ref{WorldSheetPic}). Its all-order perturbative form was bootstrapped in terms of ladder integrals with accompanying coefficients constrained by means of Steinmann relations \cite{Coronado:2018cxj}. $\mathbb{O}$ simplifies even further, in the near-null limit as we tend inter-operator squared distances to zero at the same rate $x_{i,i+1}^2 = m^2 \to 0$,
\begin{align}
\lim_{x^2_{i,i+1} \to 0} \lim_{K \to \infty} G_{4, K}/G_{4, K}^{\rm tree} = \mathbb{O}_0^2 (w_1,w_2; g)
\, .
\end{align}
Here $w_{1,2}$ are the two conformal cross ratios introduced before in Eq.\ (\ref{2pCC1}). The near-null limit of the octagon can be found in a closed form. This was largely possible thanks to its relation to Fredholm determinants \cite{Kostov:2019auq,Belitsky:2019fan} uncovered via the non-perturbative hexagonalization framework of Ref.\ \cite{Basso:2015rta,Fleury:2016ykk}. It reads \cite{Belitsky:2019fan}
\begin{align}
\label{M4dciFull}
\log\mathbb{O}_0 = -\frac{\Gamma_{\rm oct}(g)}{16}\log^2(w_1w_2)-\frac{g^2}{4}\log^2\left( 
\frac{w_1}{w_2} \right)-\frac{D(g)}{2} + O(m^2)
\, .
\end{align}
Substituting $w_1$ and $w_2$ as in Eq.\ (\ref{2pCC2}), one can see that
\begin{align}
\label{M4Octagon}
\log\mathbb{O}_0 = \log M_4
\, .
\end{align}
This observation was first made in Ref.\ \cite{Caron-Huot:2021usw}, where it was verified to four loops by comparing the sets of integrals arising in the decomposition of the octagon and the ratio function of the four W-boson amplitude. 

\subsection{Decagon}

Going to five points, the heavy correlation function of interest is 
\begin{equation}
\label{Korr51}
G_{5,K}
=
\big\langle{\rm tr} X^{2K}(x_1) \ {\rm tr}\bar{X}^K\bar{Y}^K (x_2) \ {\rm tr}Y^K\bar{Z}^K (x_3) \ {\rm tr}Z^{2K}(x_4) \ {\rm tr}\bar{Z}^K\bar{X}^K(x_5) \big\rangle
\, .
\end{equation}
We are again interested in the limit when the `frame' of the correlator becomes impassable to gluons as $K \to \infty$, as well as when all operators become nearly null pair-wise. As in the four-point case, one can now introduce an observable known as the decagon  \cite{Dectagon,Dectagon2}
\begin{align}
\lim_{x^2_{i,i+1} \to 0} \lim_{K \to \infty} G_{5, K}/G_{5, K}^{\rm tree} = \mathbb{D}_0^2 (u_1,\dots, u_5; g)
\, ,
\end{align}
shown in the right panel of Fig.\ \ref{WorldSheetPic} as an illustration. Here, $\mathbb{D}_0$ is a function of 't Hooft coupling and depends on the five conformal cross ratios introduced in Eq.\ (\ref{UsAndVs}).

Unfortunately, contrary to the four-point case, the functional space for the decagon in generic kinematics is not known; thus, it is currently impossible to bootstrap it in the same fashion as $\mathbb{O}$ \cite{Coronado:2018ypq,Coronado:2018cxj}. The hexagonalization framework is also highly inefficient for this observable, even at weak coupling, due to the extremely elaborate structure of the interaction dynamics of fundamental excitations. However, since we are interested only in the near-null kinematics, when all conformal cross-ratios are small and scale at the same rate, there is a shortcut to the sought-after analytic expressions, at least at lowest orders of perturbative series. All we need are the lowest two.

Let us introduce the so-called double-scaling (DS) limit \cite{Belitsky:2019fan} arising in the situation where 't Hooft coupling is sent to zero while all pair-wise distances become near null, such that the scaling variables
\begin{align}
t_i^2 =  g^2 \log v_{i-1} \log v_i
\, ,
\end{align}
are kept fixed. These are known as the {\rm cusp} times \cite{Olivucci:2021pss,Olivucci:2022aza}. Under these conditions, the octagon \cite{Belitsky:2019fan} and all higher polygons without internal bridges  \cite{Olivucci:2021pss,Olivucci:2022aza} become Gaussian. The near-null decagon in question scales as \cite{Olivucci:2022aza}
\begin{align}
\label{DtheStampede}
\left. \mathbb{D}_0 \right|_{\rm DS} = \exp\Big( - \sum\nolimits_{i = 1}^5 t_i^2 \Big)
\, .
\end{align}
This furnishes a powerful consistency condition on its form away from the DS-limit.

As we stated above, although integrals spanning the function-space of $\mathbb{D}_0$ are generally not known, we do know the set of loop integrals that can arise in its perturbative decomposition at one and two loops \cite{Dectagon2}. Thus, we can write
\begin{align}
\label{D0inPT}
\mathbb{D}_0 = g^2 \mathbb{D}_0^{(1)} + g^4 \mathbb{D}_0^{(2)} + \ldots
\, ,
\end{align}
with generic Ans\"{a}tze for $\mathbb{D}_0^{(1,2)}$ being
\begin{align}
\label{D02loopAnz}
\mathbb{D}_0^{(1)} = a_1 \mathbb{P}_5 \, {\rm Box}
\, , \qquad
\mathbb{D}_0^{(2)} = \mathbb{P}_5 \big[b_1 \, {\rm DBox} + b_2 \, {\rm TBox} + b_3 \, {\rm PBox} \big]
\, .
\end{align}
with some unknown numerical coefficients. All emerging integrals, with the exception of the {\sl turtle} double box ${\rm TBox}$, we have already encountered before. They are identical to Eqs.\ (\ref{BoxEq}-\ref{PBoxEq}) as we pass from momenta to dual variables $x_i$ via $p_i = x_{i,i+1}$. The TBox is defined as
\begin{align}
\mbox{TBox}
&
=
s_{12} s_{51} s_{45}
\ \ \,
\parbox[c][30mm][t]{58mm}{
\insertfig{5.7}{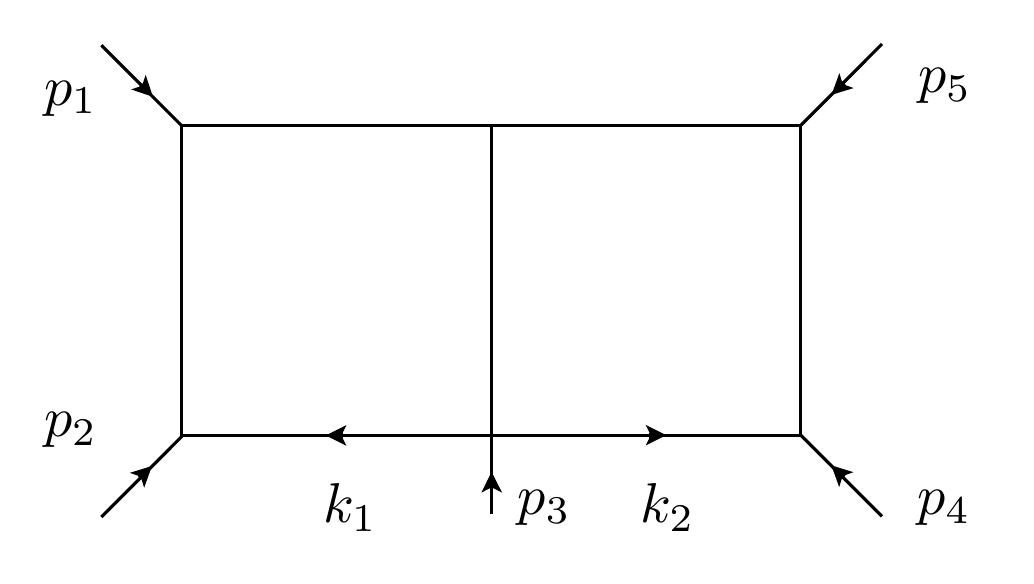}
}
\, .
\label{TBoxEq}
\end{align}
With results reported in Appendix \ref{Appendix}, we immediately deduce their asymptotic behavior in the double-scaling limit to be
\begin{align}
\label{DBandPBinStampede}
g^2 \left. {\rm Box} \right|_{\rm DS} 
&=- \frac{1}{2} \big( t_1^2 + t_5^2 \big)
\, , \\
g^4 \left. {\rm DBox} \right|_{\rm DS}
&
=\frac{1}{16} \big( t_1^2 + t_5^2 \big)^2
\nonumber\\
g^4 \left. {\rm PBox} \right|_{\rm DS}
&
=\frac{1}{8} \big( t_2^2 t_3^2+t_4^2 t_5^2 \big)
+ 
\frac{1}{4} \big( t_1^2 t_4^2+t_1^2 t_3^2 \big)
\, , \nonumber
\end{align}
such that we can reproduce the DS-limit of the decagon (\ref{DtheStampede}) even without the knowledge of the ${\rm TBox}$, though it can be found in Eq.\ (\ref{eq:doublebox_5}). The perturbative expansions at small $t_i$'s can then be matched with the following simple choice\footnote{Notice that this consideration follows word-for-word the {\sl stampede} analysis in Ref.\ \cite{Dectagon2}, however, our expressions differ. While we do agree with DS-asymptotics of the integrals in Eq.\ (\ref{DBandPBinStampede}), we disagree both with the overall normalization and relative coefficients in their representation of the near-null decagon. The main problem is that paper does not seem to have uniform conventions either for the 't Hooft coupling $g$ or for the perturbative expansions of the cusp and octagon anomalous dimensions. We failed to reproduce our expressions using theirs with a unique rescaling of 't Hooft coupling.}
\begin{align}
\label{cCoeff}
a_1=\frac{1}{2} \, , 
\quad
b_1=1 \, , \quad b_2=0 \, , \quad b_3=\frac{1}{2} \, .
\end{align}
Comparing these to the integral expansion of the five W-boson amplitude, we conclude that the two coincide with each other in the near mass-shell regime,
\begin{align}
\log\mathbb{D}_0 = \log M_5 + O(m^2)
\, .
\end{align}
This identification provides another piece of evidence for the duality relation between Coulomb branch amplitudes and heavy correlation functions.

\section{Conclusions}

In this work, we analyzed the near mass-shell limit of the Coulomb branch amplitude of five W-bosons to two-loop order in 't Hooft coupling. This was made possible by recent calculational advances, which made the previously unknown near mass-shell limit of the pentabox integral available in an analytic form. The starting point of our consideration was a higher-dimensional perspective that related scattering amplitudes in higher dimensions in theories with unbroken gauge symmetry groups to the four-dimensional ones, but with a Higgs mechanism that endows certain particles with masses. We interpreted the latter as virtualities, or off-shellness. In this manner, we found a very concise form for the five-leg amplitude. While its generic infrared structure follows that of the conformal branch of the theory, details differ significantly. First and foremost, the infrared logarithms, while being captured by a Sudakov form factor, possess unrecognizably different dependence on 't Hooft coupling compared to the massless case. While the latter is governed by the cusp, the off-shell case is driven by the octagon anomalous dimension. Second, the finite remainders vary as well. While the conformal case possesses a unique kinematical dependence on Mandelstam-like invariants accompanied by the cusp anomaly to all orders in coupling, the near mass-shell amplitude displays a less constrained form. It appears to have just {\sl two} but nevertheless independent functions of the 't Hooft coupling accompanying kinematical logarithms. While one of them is unarguably given by the octagon anomaly, the other one resembles (to the currently considered order in $g^2$) the cusp anomalous dimension. It is a question for future studies to verify this finding.

The amplitude of four W-bosons was found in earlier studies \cite{Caron-Huot:2021usw} to be equivalent in the near mass-shell limit to the `square root' of a four-point correlation function of infinitely heavy half-BPS operators, the octagon. We generalized that statement to the five-point case, finding duality of the amplitude of five W-bosons to the near-null decagon. This provides another evidence for a new duality between amplitudes and correlators, paralleling similar results in the purely massless case.

There are a number of further studies that need to be conducted to put our conjectures on a firmer foundation. The five-leg amplitude begs for consideration at three-loop order. By combining techniques rooted in tropical geometry of Feynman integrals \cite{Salvatori:2024nva,Belitsky:2025sin} and dual-conformally invariant approach to the Method of Regions \cite{Bork:2025ztu}, it is likely to be possible. The six-leg W-boson amplitude will undoubtedly be a challenge for the currently available techniques. While the set of two-loop scalar Feynman integrals defining it is now available \cite{BelitskyTBP}, their calculation will undoubtedly require new ideas. The reason being, in addition to conformal cross ratios, which are vanishing as the off-shellness goes to zero, there are three that remain oblivious to the limit in question. They stay fixed. Will this complicate the analytic evaluation of Feynman integrals? Or can they be tackled with the very same techniques as the five-leg case? It remains to be seen.

\acknowledgments
The work of L.B.\ was supported by the Basis Foundation. The work of V.S.\ was conducted under the state assignment of Lomonosov Moscow State University. 

\appendix
\section{Pentabox and other two-loop integrals}
\label{Appendix}

In this appendix, we present for completeness the results for all two-loop integrals used in our computation, see Refs. \cite{Belitsky:2025sin,Bork:2025ztu}.

For the pentabox integral \eqref{PBoxEq} we have 
\begin{multline}
\label{PB2}
    {\rm PBox}=\tfrac{1}{2} L_5 \left(L_2 L_1^2+L_3 L_4^2+2 L_2 L_3 L_1+2 L_2 L_3 L_4\right)
    +\tfrac{1}{2} \big(4 L_1 L_5+4 L_4 L_5 + 4 L_2 L_3+2 L_2 L_5
    \\
    +2 L_3 L_5 -2 L_3 L_1-2 L_2 L_4-L_1^2-L_4^2\big)\zeta _2
    +\left(L_2+L_3-2 L_5\right) \zeta_3+5 \zeta _4 +O(m^2),
\end{multline}
where now $L_i\equiv \log v_i$.  
As a crosscheck, for the symmetric point, $s_{i,i+1}=Q^2$, we find, in agreement with Ref. \cite{Bork:2022vat}, 
\begin{equation}
\label{PBintSymPoint}
{\rm PBox}\Big{|}_{s_{i,i+1}=Q^2}=3L^4+5\zeta_2L^2+5\zeta_4+O (m^2),
\end{equation}
with $L=\log (m^2/Q^2)$.

For the double box integrals \eqref{DBoxEq} we have \cite{Belitsky:2023gba,Bork:2025ztu}
\begin{multline}\label{DB}
{\rm DBox} ={\rm DBox}_1 =\left.{\rm DBox}_2\right|_{\substack{v_1\leftrightarrow v_4\\v_2\leftrightarrow v_3}} = \tfrac{1}{4} L_2^2 \left(L_3+L_1\right)^2+\tfrac{1}{2}  \Big(L_3^2+L_1^2+2 L_1 L_3\\+L_2^2+4 L_2 L_3+4 L_2 L_1\Big)\zeta _2
+\tfrac{21 }{2}\zeta _4+O(m^2)
\end{multline}
in perfect agreement with the original Davydychev-Usyukina result.
Note that this formula is symmetric with respect to $1\leftrightarrow 3,\ 4\leftrightarrow 5$. Therefore, we observe the ``magic identity'' \cite{Drummond:2006rz} ${\rm DBox}_1 =\left.{\rm DBox}_2\right|_{i \to i-1}$, which implies that $\mathbb P_5{\rm DBox}_1= \mathbb P_5{\rm DBox}_2$.
Also, as a cross-check for the symmetric point, $s_i=Q^2$, we find
\begin{eqnarray}\label{DBsym}
{\rm DBox}\Big{|}_{s_{i,i+1}=Q^2}&=&L^4+\frac{13\zeta_2}{2}L^2+\frac{21\zeta_4}{2}+O (m^2) \, .
\end{eqnarray}
Finally, for the ${\rm TBox}$ integral \eqref{TBoxEq} we find:
\begin{multline}
\label{eq:doublebox_5}
{\rm TBox} 
= 
\tfrac{1}{4} L_4 L_3 \left(2 L_5+L_3\right) \left(2 L_2+L_4\right) +\left(2 L_4 L_5+2 L_3 L_2+2 L_4 L_3+L_2 L_5\right)\zeta _2 \\
- \left(L_2+L_5\right)\zeta _3+\tfrac{31 }{4}\zeta _4
+O(m^2),
\end{multline}
which is also in agreement with symmetric point computation of \cite{Bork:2022vat}
\begin{eqnarray}
{\rm TBox}\Big{|}_{s_{i,i+1}=Q^2}&=&\frac{9}{4}L^4+7 \: \zeta_2L^2-2\zeta_3L+\frac{31}{4} \zeta_4+O(m^2).
\end{eqnarray}

\bibliographystyle{JHEP}
\bibliography{5p_ampl-corr_2loop}
\end{document}